# Stoichiometric FeTe is a Superconductor

Zi-Jie Yan[1,6], Zihao Wang[1,6], Bing Xia[1], Stephen Paolini[1], Ying-Ting Chan[2], Nikalabh Dihingia[3], Hongtao Rong[1], Pu Xiao[1], Kalana D. Halanayake[3], Jiatao Song[1], Veer Gowda[1], Danielle Reifsnyder Hickey[3,4], Weida Wu[2], Jiabin Yu[5], Peter J. Hirschfeld[5], and Cui-Zu Chang[1]

[1]Department of Physics, The Pennsylvania State University, University Park, PA 16802, USA

[2]Department of Physics and Astronomy, Rutgers University, Piscataway, NJ 08854, USA

[3]Department of Chemistry, The Pennsylvania State University, University Park, PA 16802, USA

[4]Department of Materials Science and Engineering, The Pennsylvania State University, University Park, PA 16802, USA

[5]Department of Physics, University of Florida, Gainesville, FL 32611, USA

[6]These authors contributed equally: Zi-Jie Yan and Zihao Wang

Corresponding author: cxc955@psu.edu (C.-Z. C.).

**Abstract: Iron-based superconductors are a fascinating family of materials in which multiple electronic bands and strong antiferromagnetic (AFM) correlations are key ingredients for competing ground states[1-13], including antiferromagnetism, electronic nematicity, and unconventional superconductivity. FeTe, unlike its superconducting isostructural counterpart FeSe, has long been regarded as an AFM metal sans superconductivity[14-16]. In this work, we employ molecular beam epitaxy to grow FeTe films and perform post-growth annealing under a Te flux. By performing spin-polarized scanning tunneling microscopy and spectroscopy, we demonstrate that the AFM order in as-grown FeTe films is induced by interstitial Fe atoms that disrupt the ideal 1:1 stoichiometry. Remarkably, the removal of these interstitial Fe atoms through Te annealing yields stoichiometric FeTe films that show no AFM order and instead exhibit robust superconductivity with a critical temperature of**

**~13.5K. This superconducting state is further confirmed by the observation of Cooper pair tunneling, zero electrical resistance, and the Meissner effect. Therefore, our results demonstrate that stoichiometric FeTe is inherently a superconductor, overturning a long-held view that it is an AFM metal. This work clarifies the origin of superconductivity in FeTe-based heterostructures[17-23] and demonstrates the importance of stoichiometry control in understanding the competition between AFM and superconductivity in iron-based superconductors.**

**Main text:** The discovery of superconductivity in fluorine-doped LaFeAsO in 2008(ref.[1]) marked the beginning of a new era of iron-based superconductors. Over the past two decades, iron-based superconductors have attracted intense interest due to their rich phase diagrams, unconventional pairing mechanisms, and the interplay between multiband electronic structure and strong antiferromagnetic(AFM) correlations[2-9]. Among iron-based superconductors, the iron chalcogenides have drawn particular attention owing to their structural simplicity[6-9] and their potential to host Majorana zero modes[10-13]. Similar to cuprates, the iron chalcogenides often exhibit a competition between AFM and superconductivity in their phase diagrams[4-9]. For example, FeSe is a superconductor[24], and its monolayer form on $SrTiO_3$(100) exhibits an enhanced superconducting transition temperature exceeding 60K(refs.[25-27]). In contrast, FeTe, despite being isostructural to FeSe, has long been viewed as an AFM metal without superconductivity, characterized by a bicollinear AFM ground state[14-16]. This longstanding view has shaped how FeTe is perceived in both theoretical and experimental studies[14-16], placing it outside the family of iron chalcogenide superconductors. Instead, FeTe is regarded as a parent compound for inducing superconductivity through tensile stress[28], interfacial engineering[17-23], or chemical substitution[29-31]. However, the origin of its AFM order and its relation to the superconductivity have yet to be explored.

Over the past decade, superconductivity emerges when FeTe is interfaced with various Te-based compounds[17-23], despite their differences in electronic band structure, carrier type, work function, or magnetic order. The common presence of Te in the partner layer of FeTe suggests that a Te-rich atmosphere may play a crucial role in inducing superconductivity. Meanwhile, FeTe single crystals are known to accommodate excess Fe atoms at interstitial sites, which strongly influence both magnetic and electronic behaviors[32,33]. These interstitial Fe atoms induce and stabilize the bicollinear AFM order in $Fe_{1+x}Te$ single crystals, where $x$ typically ranges from 0.02 to 0.2(refs.[34-39]). Prior studies[40-43] have shown that superconductivity in $Fe_{1+x}(Se,Te)$ is highly sensitive to $x$, which can be partially reduced through Se/Te annealing, a process referred to as stoichiometry control. To date, much effort has been made to induce superconductivity in FeTe by reacting interstitial Fe atoms with other chalcogen elements (e.g., S or Se)[29-31,34,35]. However, these approaches inevitably introduce chemical disorder and/or extrinsic dopants, making it difficult to single out the intrinsic ground state of FeTe. Therefore, whether stoichiometric FeTe is inherently an AFM metal or a superconductor has remained an open question.

In this work, we address this longstanding question by synthesizing FeTe films using molecular beam epitaxy(MBE) and precisely controlling their stoichiometry through post-growth annealing in a Te flux. Through spin-polarized scanning tunneling microscopy and spectroscopy(STM/S), we reveal that the bicollinear AFM order observed in as-grown FeTe films is stabilized by interstitial Fe atoms. As these interstitial Fe atoms are progressively removed via Te annealing, the AFM domains shrink and eventually vanish, resulting in a stoichiometric FeTe phase without AFM order. Remarkably, robust superconductivity with an onset superconducting transition temperature $T_{c,onset}$~13.5K emerges in stoichiometric FeTe films, which is confirmed by the observations of a superconducting gap and Abrikosov vortices in STM/S, a zero-resistance state in electrical transport, and a uniform Meissner effect detected by magnetic force microscopy(MFM). We perform Josephson STM/S measurements to further confirm the presence of phase-coherent Cooper pairing. Our theoretical calculations show that the interstitial Fe atoms suppress superconductivity and stabilize the bicollinear AFM order in FeTe. Together, our results unambiguously demonstrate that stoichiometric FeTe is an intrinsic superconductor rather than an AFM metal. This work upends conventional wisdom and redefines the phase diagram of iron chalcogenide superconductors.

**Absence of AFM in stoichiometric FeTe**

Figure 1a shows the MBE growth of an FeTe film, where Fe and Te elements are co-evaporated onto a SrTiO$_3$(100) substrate(Methods). The *Néel* temperature $T_N$ of as-grown FeTe films with different thicknesses ranges from 45 to 65K(refs.[17-19,44]), below which a bicollinear AFM order emerges. This bicollinear AFM order forms a magnetic unit cell(UC) of $a_{Te} \times 2a_{Te}$, doubling its structural UC of $a_{Te} \times a_{Te}$. Both UCs can be resolved by spin-polarized STM/S measurements[44-46]. The tunneling current is modulated by the relative orientation between the STM tip moment and local spin polarization(Extended Data Fig. 1a), resulting in a height contrast in STM images. Figure 1b shows a 20×20nm$^2$ STM image of an as-grown 40UC FeTe measured at a sample bias $V_{bias}$= −15mV, revealing a double-stripe pattern. This pattern shifts by $a_{Te}$ when the magnetization of the STM tip is reversed by applying opposite magnetic fields(Extended Data Fig. 1), confirming the presence of bicollinear AFM order in as-grown FeTe. Furthermore, a pair of AFM peaks (**Q**$_{AFM}$) is seen in the Fourier transform(FT) of the STM image, located at half the wavevector of the corresponding Bragg peaks (**Q**$_{Bragg}$)(Fig. 1b inset).

Next, we perform STM measurements on the same 20×20nm$^2$ area of the as-grown 40UC

FeTe at a higher $V_{bias}$=3.5V. At this bias, interstitial Fe atoms become visible, while the structural and/or magnetic contrast is not resolved(Extended Data Fig. 2). This imaging condition allows us to map the spatial distribution of interstitial Fe atoms. The as-grown FeTe film exhibits a high density of interstitial Fe atoms, with ~320 atoms detected within the 20×20nm$^2$ area (i.e., $x$=0.060)(Extended Data Fig. 3). To achieve a stoichiometric FeTe film, we anneal the as-grown 40UC FeTe under a Te flux(Fig. 1c). A stoichiometric FeTe film is achieved after 5 cycles of Te annealing treatments (denoted as Cycles I to V, Methods). Figure 1d shows a 20×20nm$^2$ STM image of the resulting stoichiometric 40UC FeTe, measured at $V_{bias}$=15mV, with no signature of bicollinear AFM order. This observation highlights the critical role of interstitial Fe in stabilizing the bicollinear AFM order in FeTe and underscores the importance of stoichiometry in determining its ground state. Moreover, before and after Te annealing, the lattice constants of FeTe films remain unchanged, as supported by the identical peak positions in X-ray diffraction spectra(Fig. 1e) and reflection high-energy electron diffraction patterns(Fig. S1). Therefore, Te annealing alters the stoichiometry of FeTe while maintaining its lattice structure. The change in stoichiometry affects the magnetic properties of FeTe.

**Correlation between AFM and interstitial Fe in FeTe**

We first examine the evolution from as-grown to stoichiometric FeTe. Figure 2a shows a 20×20nm$^2$ STM image on a 40UC FeTe after 1 cycle of Te annealing(i.e., Cycle I, Methods), measured at $V_{bias}$= −10mV. Compared to the as-grown FeTe(Fig. 1b), the area exhibiting the bicollinear AFM order with a double-stripe pattern shrinks. The newly appearing area shows only structural UC, suggesting the absence of AFM. The AFM area on the same FeTe film further shrinks after Cycle II(Fig. 2b), nearly disappears after Cycle III(Fig. 2c), and vanishes completely after Cycle IV(Extended Data Fig. 4a). After each annealing cycle, we extract the percentage of the bicollinear AFM area and plot it as a function of Te annealing cycles(Fig. 2g). We find that stoichiometric FeTe films without bicollinear AFM order and visible defects are gradually formed through successive Te annealing. This evolution is further supported by a monotonic decrease in the normalized $\mathbf{Q}_{AFM}$ peak intensity(Fig. 2h) after each annealing cycle, consistent with the shrinkage of AFM regions.

To investigate changes in the number of interstitial Fe atoms and their correlation with the bicollinear AFM order after each annealing cycle, we perform STM measurements on the same areas in Fig. 2a-c and Extended Data Fig. 4a at $V_{bias}$=3.5V. The number of interstitial Fe atoms detected within a 20×20nm$^2$ area decreases gradually(Fig. 2i): ~153 after Cycle I (i.e., $x$=0.028), ~119 after Cycle II (i.e., $x$=0.022), ~63 after Cycle III (i.e., $x$=0.012), and ~7 after Cycle IV (i.e., $x$=0.001)(Fig. 2d-f and Extended Data Fig. 4b). Next, by overlaying the STM images measured at low and high $V_{bias}$ after each annealing cycle, we find that interstitial Fe atoms in FeTe are strictly confined to the areas with bicollinear AFM order(Fig. 2d-f inset). In contrast, the areas without AFM order show no interstitial Fe. This observation indicates a strong correlation between the presence of interstitial Fe and the emergence of AFM order in FeTe, consistent with the absence of AFM in stoichiometric FeTe.

Figure 2j shows a schematic of the evolution from as-grown to stoichiometric FeTe through Te annealing. When an FeTe film is exposed to a Te-rich atmosphere at an elevated temperature, the Te atoms react with interstitial Fe atoms, forming new FeTe molecules, which grow along the edges of existing FeTe layers and/or generate small FeTe islands on top. This change in surface topography is observed in both large-scale STM and atomic force microscopy images(Extended Data Fig. 5). Besides STM, we perform cross-sectional annular dark-field scanning transmission electron microscopy(ADF-STEM) measurements on both as-grown and stoichiometric 40UC FeTe.

In the ADF-STEM images, the trilayer structure of FeTe is resolved, with the brighter dots corresponding to Te atoms due to their higher atomic number, while the dimmer dots correspond to Fe atoms. For the as-grown FeTe film, interstitial Fe atoms are observed at multiple interstitial sites within the Te layers(Fig. 2k). In contrast, the stoichiometric FeTe film shows almost no interstitial Fe atoms in the Te layers(Fig. 2l), confirming the achievement of a nearly ideal 1:1 stoichiometry throughout the entire FeTe film.

**Emergent superconductivity in stoichiometric FeTe**

We explore the electronic properties of stoichiometric FeTe films using different techniques, including STM/S, electrical transport, and MFM. Figure 3a shows a typical differential conductance $dI/dV$ spectrum of the stoichiometric 40UC FeTe measured at $T$=310mK. A superconducting gap $\Delta$~4.4meV is observed, indicating the emergence of superconductivity. More discussion of the $\Delta$ value in stoichiometric FeTe can be found in Supplementary Information. As $T$ increases, $\Delta$ gradually decreases and eventually disappears at $T$~13.5K(Fig. 3b). Upon applying a perpendicular external magnetic field $\mu_0 H$, a hexagonal lattice of Abrikosov vortices is observed(Fig. 3c and Extended Data Fig. 6). Figure 3d shows a color plot of $dI/dV$ spectra measured across an Abrikosov vortex at $\mu_0 H$=8T. Several bound states at asymmetric $V_{bias}$ are observed within the superconducting gap at the vortex core(Fig. 3e), different from the symmetric Caroli-de Gennes-Matricon bound states[47-49] commonly observed in iron-based superconductors[10-13,50-54].

To further confirm that stoichiometric FeTe is a superconductor, we perform Josephson STM/S measurements on the stoichiometric 40UC FeTe using a superconducting Nb tip(Extended Data Fig. 7). When a Nb tip approaches the surface of a superconductor, a superconductor-insulator-superconductor(S-I-S) Josephson junction forms between the Nb tip and the sample surface. The tunneling barrier is characterized by the normal state junction resistance $R_N$, which can be tuned by varying the tip-to-sample distance $D$(Fig. 3f). By gradually reducing $R_N$ to ~0.80MΩ, we observe a double kink feature in the $I$-$V_{bias}$ curve(Fig. 3g) and a pronounced zero-bias peak in the corresponding d$I$/d$V$ spectrum(Fig. 3h), indicating Cooper-pair tunneling[55-57]. As the Nb tip is retracted from the surface of the stoichiometric FeTe film, the Josephson tunneling signal gradually weakens and nearly disappears at $R_N$~3.78MΩ.

Next, we perform electrical transport and MFM measurements to demonstrate the zero-resistance state and the Meissner effect of the stoichiometric FeTe films, respectively. Figure 4a shows the $T$-dependent sheet longitudinal resistance $R_{xx}$ of both as-grown and stoichiometric 40UC FeTe. For as-grown FeTe, a peak feature appears at $T$~53K, corresponding to its $T_N$, consistent with prior studies[22-24,44]. Therefore, the as-grown FeTe film is a nonsuperconducting AFM metal. After each Te annealing cycle, the $T_N$ hump feature gradually weakens, indicating suppression of the bicollinear AFM order. Simultaneously, a superconducting phase transition progressively establishes in corresponding $R_{xx}$-$T$ curves(Extended Data Fig. 8). These observations agree well with our STM/S results(Fig. 2g and Extended Data Fig. 9). For stoichiometric FeTe, the $T_N$ peak feature is indiscernible and a superconducting phase transition occurs at $T$~13.5K, with a zero-resistance state observed at $T_{c,0}$~12.0 K, consistent with our STM/S measurements(Fig. 3b).

Besides the zero resistance state, we perform MFM measurements on the stoichiometric 40UC FeTe to detect the Meissner effect, i.e., the spontaneous expulsion of magnetic fields by supercurrent(Fig. 4b)[58]. The resulting repulsive interaction between the magnetic MFM tip and the superconductor is detected as a positive shift in the cantilever's resonance frequency $df$ (ref.[59]). This repulsive interaction becomes stronger as the superfluid density increases upon cooling. Figure 4c shows the $df$-$T$ curves measured over an 8×8 grid within an 11×11μm$^2$ area(Fig. 4d-i).

Each semi-transparent gray line represents an averaged $df$-$T$ curve within one grid cell, while the black line shows the average $df$ across all 64 grid cells. All curves exhibit a sharp upturn at $T$~11.8K, marking the onset of the Meissner effect, followed by a monotonic increase at lower $T$. This pronounced $T$-dependent increase in $df$ reflects a strong Meissner response, confirming the emergence of macroscopic phase coherence in the superconducting state. For $T \geq 11.8$K, the signal remains flat and close to zero. We note that $R_{xx}$ drops to zero at the same temperature where the $df$ signal begins to rise(Fig. 4c), indicating the formation of a uniform superconducting state across the whole film. Indeed, the small variations among all grid-point curves demonstrate the uniformity of superconductivity within the scanning area. Furthermore, the observation of the Meissner effect at random locations separated by hundreds of micrometers also demonstrates that superconductivity is uniform over macroscopic length scales in our stoichiometric FeTe.

**Mechanism for stabilizing AFM order in FeTe**

Based on the $x$ values determined by STM and the $T_c$ and $T_N$ values extracted from the $R_{xx}$-$T$ curves, we establish a phase diagram of $Fe_{1+x}Te$(Fig. 4j). As $x$ increases, the ground state of $Fe_{1+x}Te$ evolves from a superconducting state for $x \leq 0.012$ to an AFM metal for $x \geq 0.022$. For $0.012 < x < 0.022$, the superconducting and AFM areas coexist in FeTe films. Next, we theoretically study the role of interstitial Fe in quenching superconductivity and stabilizing the bicollinear AFM order in FeTe. We build a 2D tight-binding model for the Fe $d$ orbitals[60,61] and include an onsite repulsive interaction, an attractive interaction for $s_{\pm}$ pairing[62], and a small next-next-nearest neighbor spin-spin correlation term $J_3$ that helps stabilize the bicollinear AFM order[14]. Interstitial Fe atoms are simulated by impurities with a local chemical potential shift and magnetic moment (Methods).

On a 24×24 lattice, we find no AFM order in the absence of impurities. As the number of impurities $N$ increases to intermediate impurity concentrations (e.g., $N=7$ and 14), a bicollinear AFM [i.e., $\left(\frac{\pi}{2},\frac{\pi}{2}\right)$] order emerges. At high impurity concentrations (e.g., $N=21$, 28, and 35), two orthogonal bicollinear AFM [i.e., coexisting $\left(\frac{\pi}{2},\frac{\pi}{2}\right)$ and $\left(\frac{\pi}{2},-\frac{\pi}{2}\right)$] orders are present(Extended Data Fig. 10 and Fig. S20). This trend is consistent with our experimental observations(Figs. 1, 2, and S2) as well as a prior study[45]. Moreover, the average Cooper pairing strength $|\Delta|$ is 4~5meV for $N=0$, which is close to $\Delta$~4.4meV observed in our STS measurements on stoichiometric FeTe(Fig. 3a). $|\Delta|$ eventually drops to nearly zero for $N\geq 6$(Fig. S21), consistent with the disappearance of superconductivity at high interstitial Fe concentrations(Extended Data Figs. 8 and 9).

**Discussion and outlook**

Our newly-established phase diagram of $Fe_{1+x}Te$(Fig. 4j) reveals that in the absence of interstitial Fe, stochiometric FeTe is a robust superconductor. This finding redefines the role of FeTe within the iron-based superconductors and highlights the dual roles of multiband electronic structure and strong AFM correlations in stabilizing unconventional superconductivity across other families of quantum materials[9]. Our work resolves a major debate regarding the origin of superconductivity in iron chalcogenides[8,9], demonstrating that superconductivity in Fe(Se,Te) is predominantly mediated by spin, rather than nematic[63,64], fluctuations. More broadly, our results reveal how stoichiometric disorder can fundamentally reshape the landscape of closely competing phases in correlated systems. As one of the simplest superconducting iron chalcogenides, FeTe offers a clean platform where, instead of a single magnetic ground state[6], varying degrees of disorder stabilize multiple competing orders. Similar phenomena are likely to be present in other correlated materials, where hidden superconducting states or competing magnetic orders remain concealed until the disorder is removed or carefully controlled.

To summarize, we demonstrate that stoichiometric FeTe is an intrinsic superconductor with $T_{c,onset}$~13.5K, overturning its long-standing classification as a nonsuperconducting AFM metal. By removing interstitial Fe atoms, we reveal the clean-limit superconducting ground state and show that the bicollinear AFM order is a disorder-stabilized phase. Stoichiometric FeTe offers a simple and clean platform for investigating unconventional pairing mechanisms in iron-based superconductors. Our findings redefine the phase diagram of iron chalcogenides and provide a broadly applicable strategy for uncovering and stabilizing hidden superconducting states in other correlated materials.

## Methods

### MBE growth and Te annealing treatments

FeTe films used in this work are grown in two commercial MBE chambers [1 Lab 10 from ScientaOmicron and 1 from Unisoku]. Each MBE chamber has a vacuum better than $3 \times 10^{-10}$ mbar. Both metallic 0.5% Nb-doped $SrTiO_3$(100) and insulating $SrTiO_3$(100) substrates are used for the MBE growth of FeTe films. FeTe films grown on metallic $SrTiO_3$(100) are used in STM/S measurements, while those grown on insulating $SrTiO_3$(100) are used in ex situ XRD, STEM, electrical transport, and MFM measurements. Before MBE growth, all $SrTiO_3$(100) substrates are first soaked in hot deionized water (~80 °C) for 2 hours, then immersed in a ~4.5% HCl solution for 2 hours, and finally annealed at ~974 °C for 3 hours in a tube furnace with flowing oxygen. These treatments passivate and reconstruct the $SrTiO_3$(100) surface, making it suitable for the MBE growth of FeTe films. These heat-treated $SrTiO_3$(100) substrates are loaded into the MBE chambers and outgassed at ~600 °C for 1 hour before the MBE growth. High-purity Fe (99.995%) and Te (99.9999%) are co-evaporated from Knudsen effusion cells. The substrate temperature is kept at ~330 °C during the MBE growth. The growth rate is ~0.2 UC per minute.

To remove excess Fe atoms, the as-grown FeTe films are annealed under a Te flux at ~280 °C. In our experiments, each Te annealing treatment cycle lasts 5 minutes, and stoichiometric FeTe films are achieved after 5 cycles. These 5 cycles are referred to as Cycles I to V in the main text. The MBE growth and Te annealing treatments are monitored using reflection high-energy electron diffraction (RHEED) patterns (Fig. S1). For STEM and XRD measurements, a 10 nm-thick Te capping layer is deposited onto the FeTe films at room temperature to prevent oxidation. Unless otherwise specified, no capping layer is involved for the samples used in other ex situ measurements. To obtain more data points for the phase diagram of $Fe_{1+x}Te$ in Fig. 4j, we treat more as-grown 40 UC FeTe films with fractional Te annealing cycles and perform systematic STM/S and electrical transport measurements.

**STM/S measurements**

The in situ STM/S measurements are performed in a Unisoku 1300 system with a base vacuum better than $3 \times 10^{-10}$ mbar, which is connected to one of the two commercial MBE chambers (Unisoku) mentioned above. The system incorporates a single-shot $^3$He cryostat to achieve a base temperature of ~310 mK. The maximum magnetic field of the system is ~11 T. Unless otherwise specified, polycrystalline PtIr tips are used in our STM/S measurements. Before STM/S measurements on FeTe films, the PtIr tips are regularly conditioned on an MBE-grown Ag film to ensure clean and stable tunneling characteristics. To prepare a magnetic tip for spin-polarized STM/S measurements, the PtIr tip is gently indented into the FeTe surface. A few Fe atoms attached to the apex of the PtIr tip make the tip spin-polarized. A similar method has been employed in prior studies[44,45,65,66]. The d$I$/d$V$ spectra are obtained using the standard lock-in method by applying a small a.c. modulation voltage $V_{mod}$ at a frequency of 987.5 Hz and a d.c. bias voltage ($V_{bias}$). All STM images are processed with WSxM 5.0 software[67]. The Lawler–Fujita drift-correction algorithm[68] is applied for achieving FT images to eliminate drift artifacts.

**Josephson STM/S measurements**

In our Josephson STM/S measurements, polycrystalline superconducting Nb tips are used to form superconductor-insulator-superconductor (S-I-S) junctions. These superconducting Nb tips are made from 0.25 mm Nb wires via electrochemical etching in concentrated HCl solution using an a.c. voltage[69]. Before Josephson STM/S measurements on FeTe films, the Nb tips are first cleaned by electron-beam heating to remove surface oxides, and then gently conditioned on an MBE-grown Ag film to sharpen the apex to ~4 nm. The superconducting gap of our Nb tips is ~1.48 meV (Extended Data Fig. 7c and Fig. S8a-c). The normal state junction resistance $R_N$ ranges from 0.80 to 3.78 MΩ for STS measurements on the stoichiometric 40 UC FeTe (Fig. 3g,h). Given the superconducting gap sizes of the Nb tip ($\Delta_{tip}$ = 1.48 meV) and the sample ($\Delta$ = 4.4 meV), the Josephson coupling energy $E_J$ is estimated to be in the range of 2.23~10.53 μeV (Supplementary Information), which is much smaller than the thermal energy $E_T = k_B T$ ~ 26.7 μeV at $T$ = 310 mK. This disparity confirms that our Josephson STM/S measurements are conducted in the phase diffusive regime[55,57,70,71]. In the measurements, the normal state resistance, $R_N$, is calculated from the tunneling current, $I_t$, at $V_{bias}$ = 15 mV using the equation $R_N$ = (15 mV)/$I_t$.

**MFM measurements**

The MFM measurements are performed using a home-built $^3$He atomic force microscopy system with commercial piezoresistive cantilevers (spring constant $k$ ~ 3 N/m and resonance frequency $f$ ~ 41 kHz). The tips are coated with a ~100 nm thick Co layer by using magnetron sputtering. The MFM data is extracted by Nanonis SPM Controllers (SPECS) with a phase-locked loop. The MFM signal, i.e., the resonant frequency shift of the cantilever ($df$), is proportional to the out-of-plane stray field gradient generated by the sample. By using the constant-height mode, MFM images are acquired on a scanning plane ~200 nm (Figs. 4 and S11) and 1 μm (Fig. S12) above the surface. All MFM images are measured at zero magnetic field, and the background component of *df* due to slight temperature variation is subtracted.

**Electrical transport measurements**

The FeTe films grown on 2 mm × 10 mm heat-treated SrTiO$_3$ (100) substrates are scratched into a Hall bar geometry using a computer-controlled motorized probe station. The effective area of the Hall bar device is ~1 mm × 0.5 mm. The electrical contacts are made by pressing indium spheres on the Hall bar. The electrical transport measurements are conducted using a Physical Property Measurement System (PPMS, Quantum Design DynaCool, 1.7 K, 9 T). The excitation current is 1 μA in all $R_{xx}$-$T$ measurements. To minimize oxidation, all samples are measured within 30 minutes after being taken out from the MBE chamber. A superconducting phase transition is consistently observed across a series of stoichiometric FeTe films with varying thicknesses (Fig. S10).

**X-ray diffraction (XRD) measurements**

The high-resolution XRD measurements are carried out at room temperature using a Malvern Panalytical X'Pert$^3$ materials research diffractometer (MRD) with Cu-$K_{\alpha 1}$ radiation (wavelength λ ~ 1.5405980 Å) in the θ-2θ geometry. The diffraction peaks are indexed according to the tetragonal crystal structure of FeTe (space group: *P*4/*nmm*), confirming the high crystallinity and phase uniformity of the as-grown and stoichiometric FeTe films.

**ADF-STEM measurements**

The FeTe samples for our cross-sectional ADF-STEM measurements are prepared using FEI

Helios 660 Nanolab and FEI Scios 2 focused ion beam (FIB) systems. To prevent damage during the FIB process, a ~400 nm thick amorphous carbon layer is deposited onto the sample surface using a Leica EM ACE200 sputter coater. Aberration-corrected ADF-STEM measurements are performed using an FEI Titan³ G2 S/TEM operated at an accelerating voltage of 300 kV. All ADF-STEM images are obtained using a probe convergence angle of ~25.3 mrad, a probe current of 80~85 pA, and collection angles of 42~244 mrad. All ADF-STEM images are bandpass filtered to (2,40) pixels using ImageJ.

**Theoretical calculations**

To understand the role of interstitial Fe atoms for stabilizing the bicollinear AFM order in FeTe, we construct a tight-binding model for FeTe based on its crystal structure in the *P*4/*nmm* space group. First, we perform DFT calculations for nonmagnetic stoichiometric FeTe using the Perdew-Burke-Ernzerhof exchange-correlation functional and experimentally determined structural parameters: $a = b = a_{Te} = 3.862$ Å, $c = 6.262$ Å, and an Fe-Te plane separation distance of 1.712 Å. The resulting DFT band structure is shown in Fig. S18c. At the $k_z = 0$ plane, the bands near the Fermi surface are predominantly composed of Fe $d$ orbitals (Fig. S18c), which motivates us to construct a two-dimensional (2D) tight-binding model for the $k_z = 0$ plane with Fe $d$ orbitals as the basis. The model is first obtained by Wannierizing the low-energy bands, then unfolded[60,61] onto a square lattice with one Fe atom per UC, i.e., lattice constant $a_{Fe} = \frac{1}{\sqrt{2}} a_{Te}$ (Fig. S18b), and finally truncated to have only the hopping terms with distances up to $2\sqrt{2}a_{Fe}$. Our 2D model accurately captures the low-energy physics of an individual Fe layer in our FeTe films, and we therefore focus on this 2D model in our correlated calculations.

To study magnetic interactions, we add an onsite repulsive Hubbard-Kanamori interaction[61,62,72]:

$$H_{int,onsite} = \frac{U}{2}\sum_{R,\alpha,\sigma}\hat{n}_{R\alpha\sigma}\hat{n}_{R\alpha\bar{\sigma}} + \frac{2U'-J}{4}\sum_{R,\alpha\neq\beta,\sigma,\sigma'}\hat{n}_{R\alpha\sigma}\hat{n}_{R\beta\sigma'} - 4J\sum_{R,\alpha\neq\beta}\hat{S}_{R\alpha}\cdot\hat{S}_{R\beta}$$
$$+ \frac{J'}{2}\sum_{R,\alpha\neq\beta,\sigma}c^{\dagger}_{R\alpha\sigma}c^{\dagger}_{R\alpha\bar{\sigma}}c_{R\beta\bar{\sigma}}c_{R\beta\sigma} \qquad (1)$$

where $c^{\dagger}_{R\alpha\sigma}$ is the creation operator for the electron, $\bm{R} = l_1\bm{a}_1 + l_2\bm{a}_2$ is the lattice vector with $\bm{a}_1 = \left(\frac{1}{\sqrt{2}},\frac{1}{\sqrt{2}}\right)a_{Fe}$ and $\bm{a}_2 = \left(-\frac{1}{\sqrt{2}},\frac{1}{\sqrt{2}}\right)a_{Fe}$ two basis lattice vectors, $\alpha$ and $\beta$ label the $d$ orbitals of Fe atoms, $\sigma$ labels the spin, $\bar{\sigma}$ is always opposite to $\sigma$, and $\hat{n}_{R\alpha\sigma} = c^{\dagger}_{R\alpha\sigma}c_{R\alpha\sigma}$ is the electron number operator. $U$ is the onsite Hubbard interaction for each orbital, $U'$ represents the Coulomb repulsion between electrons in different orbitals, $J$ is the Hund's coupling, $J'$ is the pairing hopping energy, and $\hat{S}_{R\alpha,i} = \frac{1}{2}\sum_{\sigma,\sigma'}c^{\dagger}_{R\alpha\sigma}c_{R\alpha\sigma'}[s_i]_{\sigma\sigma'}$ with $i = x, y, z$ and $s_i$ is the $i^{th}$ Pauli matrix. We neglect spin-orbit coupling, thereby preserving the spin-rotational invariance, and choose $U' = U - 2J$, $J = U/4$, and $J' = J$. Besides these onsite terms, we also phenomenologically add a next-next-nearest neighbor spin-spin interaction:

$$H_{Int,J_3} = 4J_3\sum_{R,\alpha,\delta}\hat{S}_{R\alpha}\cdot\hat{S}_{R+\delta\alpha} \qquad (2)$$

where $J_3 = 0.1U$ and $\bm{\delta} = \pm 2\bm{a}_1, \pm 2\bm{a}_2$. This $J_3$ term ensures that the homogeneous magnetic order stabilized at large repulsive interactions in the clean limit[14] corresponds to the double-stripe pattern observed in our experiments. We further add an attractive interaction for the $s_{\pm}$-wave pairing[62]:

$$H_{pairing} = -\sum_{R,\alpha,\delta}\frac{\Gamma_\alpha}{4}\left(c^{\dagger}_{R\alpha\uparrow}c^{\dagger}_{R+\delta\alpha\downarrow} - c^{\dagger}_{R\alpha\downarrow}c^{\dagger}_{R+\delta\alpha\uparrow}\right)\left(c_{R+\delta\alpha\downarrow}c_{R\alpha\uparrow} - c_{R+\delta\alpha\uparrow}c_{R\alpha\downarrow}\right) \qquad (3)$$

where $\Gamma_{Eg} = 0.5\Gamma_{t2g}$ (ref.[62]). Finally, we include the chemical potential terms $\mu$, a nonmagnetic impurity term $V_R$, and a magnetic impurity term $\bm{S}_R$, resulting in the full Hamiltonian:

$$H = H_0 + \sum_{R,\alpha,\sigma}(-\mu + V_R)\,\hat{n}_{R\alpha\sigma} + \sum_{R,\alpha} S_R \cdot \hat{S}_{R\alpha} + H_{int,onsite} + H_{Int,J_3} + H_{pairing} \quad (4)$$

where $H_0$ is the 2D tight-binding model (Supplementary Information).

We solve the Hamiltonian self-consistently within the mean-field approximation. Following ref.[45], we assume that our ground state always preserves a fixed spin direction, i.e., the magnetic order must be collinear. Without loss of generality, we take the preserved spin direction to be along the $z$-axis. We choose $\Gamma_{t2g} = 0.45$ eV. An interstitial Fe at $R$ is modeled as an impurity with $V_R$ and $S_R$ = (0, 0, $2S_R$). The position $R$ and the sign of $S_R$ are chosen randomly (Extended Data Fig. 10 and Fig. S20). The lattice vector is chosen to be $R = l_1 a_1 + l_2 a_2$ with $l_1, l_2 = 0, 1, 2, \ldots, L-1$, and we perform our calculations on a lattice with $L = 24$ with periodic boundary condition.

In particular, we focus on two order parameters, the superconducting pairing $\Delta_{R\delta\alpha}$ and the magnetization $m_R$, defined as follows:

$$\Delta_{R\delta\alpha} = -\frac{\Gamma_\alpha}{2}\langle c_{R+\delta\alpha\downarrow} c_{R\alpha\uparrow} - c_{R+\delta\alpha\uparrow} c_{R\alpha\downarrow}\rangle \quad (5)$$

and

$$m_R = \sum_\alpha \langle n_{R\alpha\uparrow} - n_{R\alpha\downarrow}\rangle \quad (6)$$

In addition, the FT of $m_R$ is used to identify any potential magnetic order and defined as

$$m_q = \frac{1}{\sqrt{L^2}}\sum_R m_R e^{iR\cdot q} \quad (7)$$

where $q$ is the linear combination of the reciprocal lattice basis vectors $b_1$ and $b_2$, corresponding to our tight-binding model. Note that the emergence of pronounced peaks at $q = \pm(b_1+b_2)/4$ or $\pm(b_1-$

$b_2)/4$ [i.e., $\left(\frac{\pi}{2}, \frac{\pi}{2}\right)$ or $\left(\frac{\pi}{2}, -\frac{\pi}{2}\right)$ order, respectively] signals the development of bicollinear magnetic order in FeTe.

**Acknowledgments:** We thank M. H. W. Chan, Y. Ge, A. J. Grutter, L. Y. Kong, C. X. Liu, Z. Q. Mao, N. Samarth, Z. Y. Wang, X. X. Wu, X. D. Xu, B. H. Yan, and J. Zhu for helpful discussions. This work is primarily supported by the DOE grant (DE-SC0023113), including the MBE growth and STM/S measurements. The atomic force microscopy and XRD measurements are supported by the ONR Award (N000142412133). The electrical transport measurements are supported by the ARO award (W911NF2210159) and the NSF grant (DMR-2241327). The STEM measurements are supported by the Penn State MRSEC for Nanoscale Science (DMR-2011839). The MFM measurements are supported by the DOE grant (DE-SC0018153). JY acknowledges the support from the startup funds at University of Florida. PJH acknowledges the support from the NSF grant (DMR-2231821). CZC acknowledges the support from the Gordon and Betty Moore Foundation's EPiQS Initiative (GBMF9063 to C.-Z. C).

**Author contributions:** CZC conceived and designed the experiment. ZJY, ZW, BX, SP, HR, and CZC performed the MBE growth. ZJY, HR, PX, and CZC conducted electrical transport measurements. ZW, BX, SP, PX, JS, VG, and CZC performed all STM/S measurements. ZJY, PX, and CZC performed the atomic force microscopy and XRD measurements. YTC and WW performed MFM measurements. ND, KDH, and DRH performed STEM measurements. JY and PJH provided theoretical support. ZJY, WW, JY, PJH, and CZC analyzed the data and wrote the manuscript with input from all authors.

**Competing interests:** The authors declare no competing financial interests.

**Data availability:** The data that support the findings of this article are openly available[73].

**Figures and figure captions:**

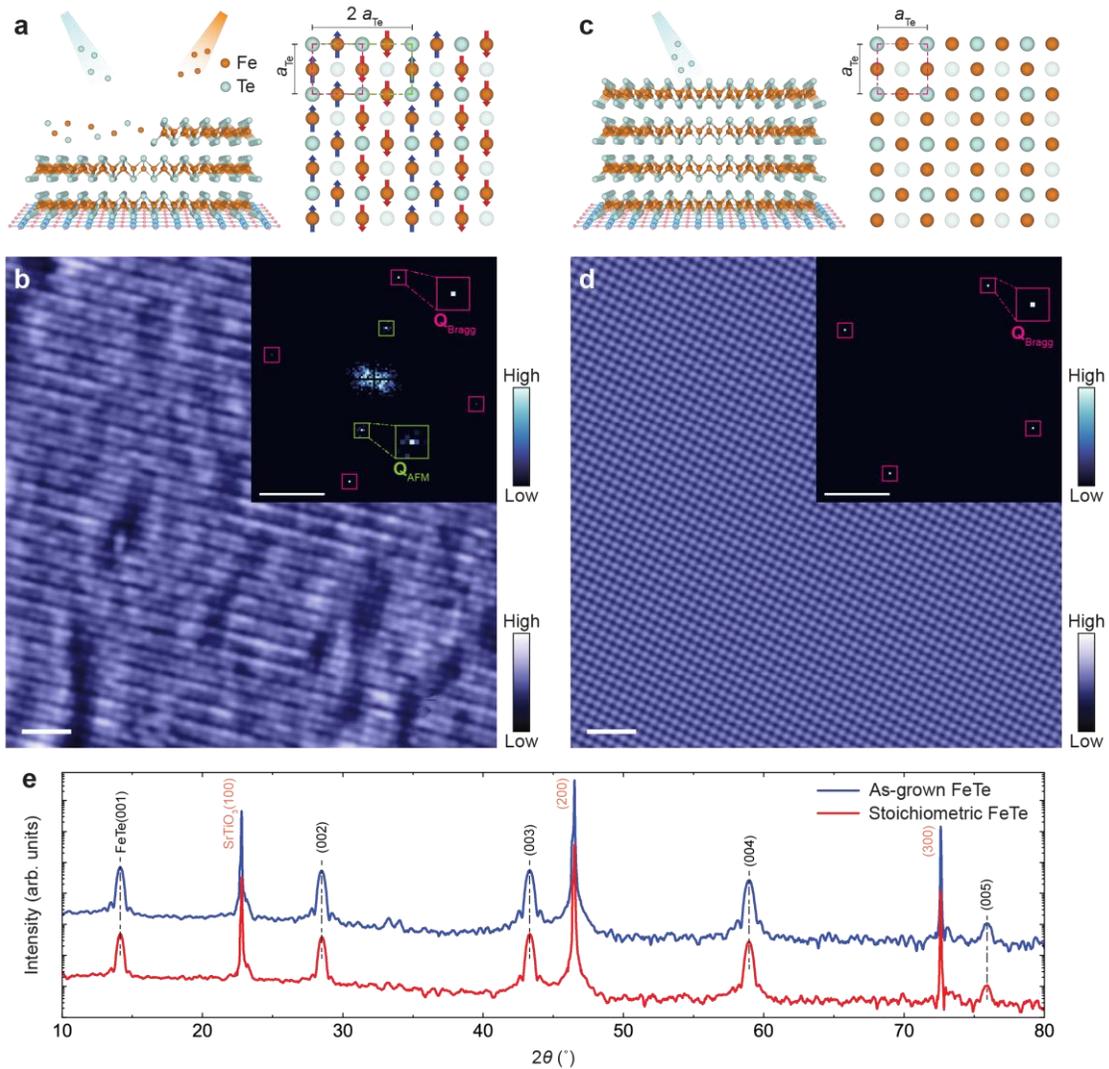

**Fig. 1| MBE-grown FeTe films before and after Te annealing treatments. a**, Left: Schematic of the MBE growth of an FeTe film. Right: Top view of the lattice structure and bicollinear AFM order of an as-grown FeTe film. The red (green) dashed rectangular represents the structural (magnetic) UC. The red and blue arrows represent the magnetic moments of the Fe atoms. **b**, STM image (20 × 20 nm$^2$) of an as-grown 40 UC FeTe film (setpoint bias $V_s = -15$ mV, setpoint current $I_s = 5$ nA, and $T = 4.2$ K). The double-stripe pattern represents the bicollinear AFM order. Inset: Fourier transform (FT) image. The red squares mark the Bragg peaks ($\mathbf{Q}_{Bragg}$) associated with the structural UC, and the green squares mark a pair of AFM peaks ($\mathbf{Q}_{AFM} = 1/2\mathbf{Q}_{Bragg}$) associated with the magnetic UC. **c**, Left: Schematic of the Te annealing treatments on the as-grown FeTe film. Right: Top view of the lattice structure of a stoichiometric FeTe film. **d**, STM image (20 × 20 nm$^2$) of a stoichiometric 40 UC FeTe film after 5-cycle Te annealing treatments ($V_s = 15$ mV, $I_s = 2$ nA, and $T = 310$ mK). Inset: FT image. The red squares mark the Bragg peaks ($\mathbf{Q}_{Bragg}$) associated with the structural UC. **e**, XRD spectra of as-grown (blue) and stoichiometric (red) 40 UC FeTe films. Scale bars: 2 nm (**b, d**); 1 Å$^{-1}$ (**b, d**, inset). The STM images in (**b, d**) are acquired using a spin-polarized tip.

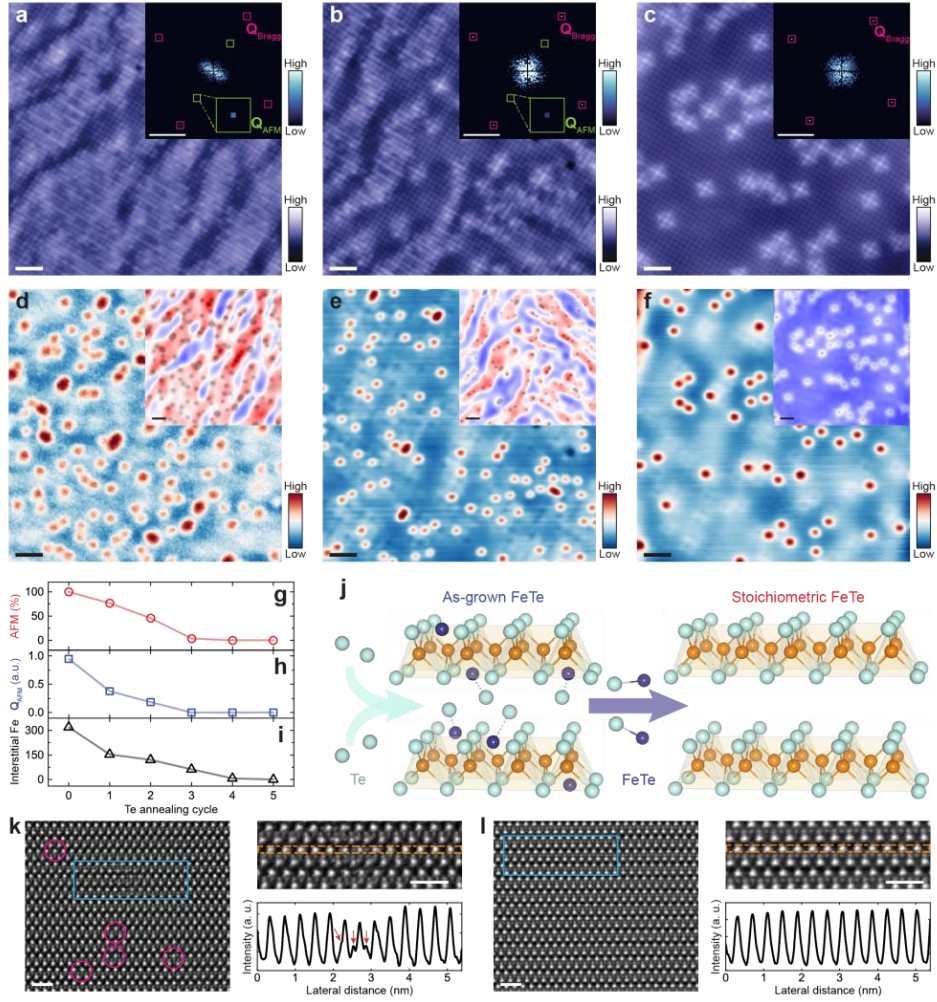

**Fig. 2| Te annealing treatments on as-grown FeTe films. a-c**, STM images (20 × 20 nm$^2$) of the 40 UC FeTe film after 1-cycle (**a**) ($V_s = -10$ mV and $I_s = 0.5$ nA), 2-cycle (**b**) ($V_s = -100$ mV and $I_s = 1$ nA), and 3-cycle (**c**) ($V_s = -50$ mV and $I_s = 5$ nA) Te annealing treatments. Inset: FT images. **d-f**, STM images (20 × 20 nm$^2$) of the same region in (**a-c**) measured using a higher $V_s$. $V_s = 3.5$ V and $I_s = 10$ pA (**d**); $V_s = 3.5$ V, $I_s = 1$ nA (**e, f**). The red spots in (**d-f**) represent the interstitial Fe atoms. Inset: The corresponding overlay of the STM images in (**a-c**) and (**d-f**). The red (blue) areas in (**d-f**, inset) represent the AFM (non-AFM) region shown in (**a-c**). The black spots in (**d-f**, inset) represent the interstitial Fe atoms in (**d-f**). **g-i**, The percentage of AFM region (**g**), intensity of the $Q_{AFM}$ peak (**h**) (normalized by the intensity of the corresponding $Q_{Bragg}$ peak), and the number of interstitial Fe atoms (**i**) as a function of the Te annealing cycle number. **j**, Schematic of the evolution from as-grown to stoichiometric FeTe films through Te annealing treatments. **k, l**, Left: Cross-sectional ADF-STEM images of as-grown (**k**) and stoichiometric (**l**) 40 UC FeTe films. The red circles in (**k**) highlight the regions with interstitial Fe atoms. Right top: Enlarged images of the regions marked by the blue boxes. Right bottom: Line profiles of the areas marked by the orange dashed boxes. The red arrows in (**k**) mark additional peaks in the line profile, corresponding to the interstitial Fe atoms in the as-grown FeTe film. Scale bars: 2 nm (**a-f**); 1 Å$^{-1}$ (**a-c**, inset); 2 nm (**d-f**, inset); 1 nm (**k, l**). The STM images in (**a-f**) are acquired using a spin-polarized tip at $T = 4.2$ K.

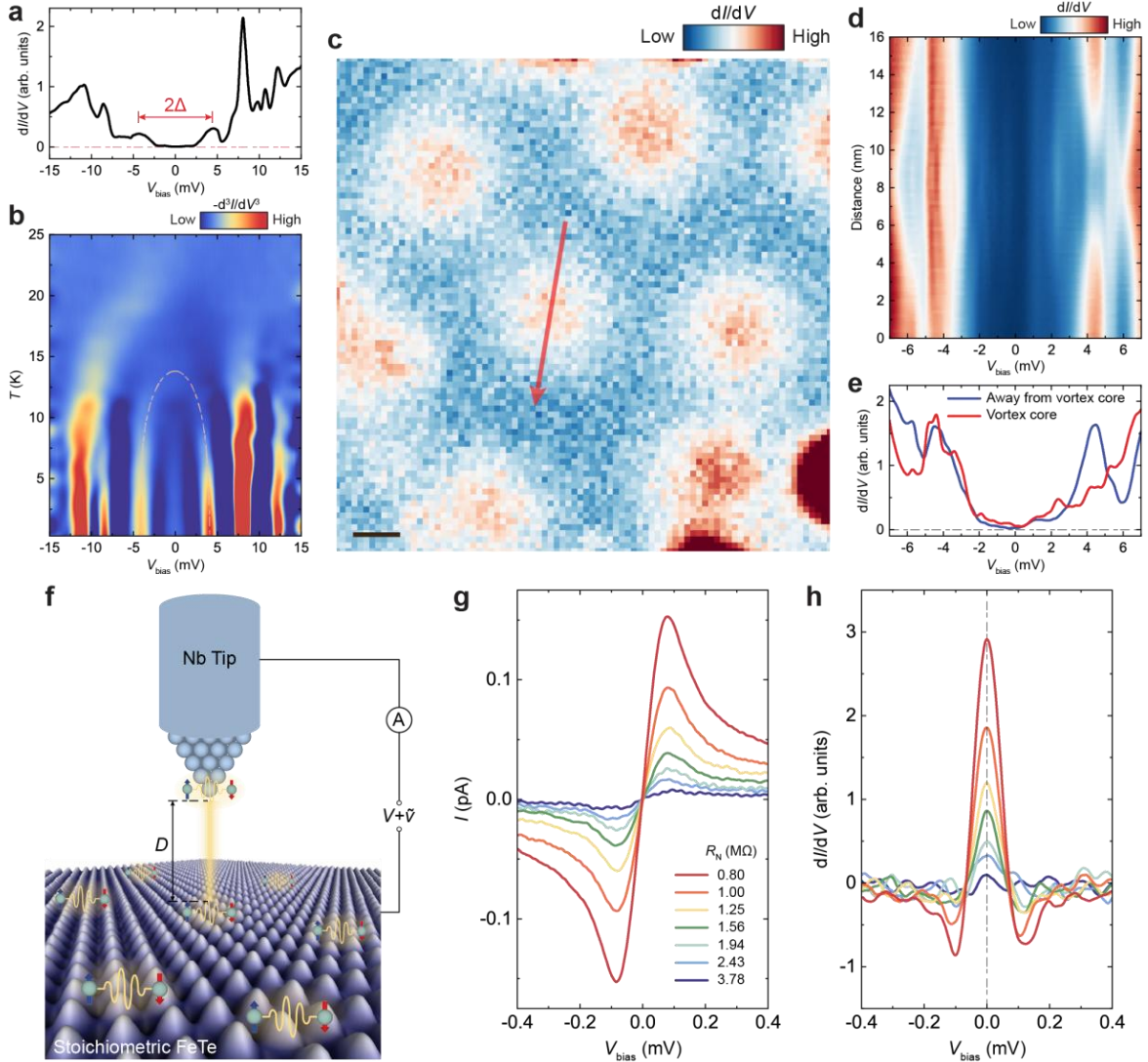

**Fig. 3| Superconductivity in stoichiometric FeTe films. a**, Typical d$I$/d$V$ spectrum ($V_s$ = 15 mV, $I_s$ = 2 nA, and modulation voltage $V_{mod}$ = 0.1 mV) of a stoichiometric 40 UC FeTe film. The superconducting gap size Δ is ~4.4 meV, determined by one half of the distance between coherence peaks. **b**, Color plot of the negative second-order derivative of the d$I$/d$V$ spectra (i.e., $-d^3I/dV^3$) at different $T$ ($V_s$ = 15 mV, $I_s$ = 2 nA, and $V_{mod}$ = 0.1 mV). The dashed line indicates the coherence peaks of the superconducting gap for reference. **c**, The Abrikosov vortices measured at $\mu_0H$ = 8 T (42 × 42 nm$^2$, $V_s$ = 15 mV, $I_s$ = 1 nA, $V_{mod}$ = 2 mV, and sample bias $V_{bias}$ = 0 mV). **d**, Color plot of d$I$/d$V$ spectra along the red arrow in (**c**) ($V_s$ = 7 mV, $I_s$ = 2 nA, and $V_{mod}$ = 0.1 mV). **e**, d$I$/d$V$ spectra at vortex core (red) and away from vortex core (blue) extracted from (**d**). **f**, Schematic of Josephson STM/S measurements on a stoichiometric FeTe film. The tunneling barrier resistance is modulated by the tip-to-sample distance $D$. **g, h,** $I$-$V_{bias}$ curves (**g**) and corresponding d$I$/d$V$ spectra (**h**) measured at different normal state junction resistance $R_N$ by gradually reducing $D$ ($D_{offset}$ from 0 pm to 70 pm) ($V_s$ = 15 mV, $I_s$ = 4 nA, and $V_{mod}$ = 0.05 mV). Josephson tunneling signals are observed at $V_{bias}$ = 0 mV. Scale bar: 4 nm (**c**). The data in (**a-e**) are measured using a normal PtIr tip. The data in (**g, h**) are measured using a superconducting Nb tip. All STM/S measurements are performed at $T$ = 310 mK, except for the $T$-dependent data in (**b**).

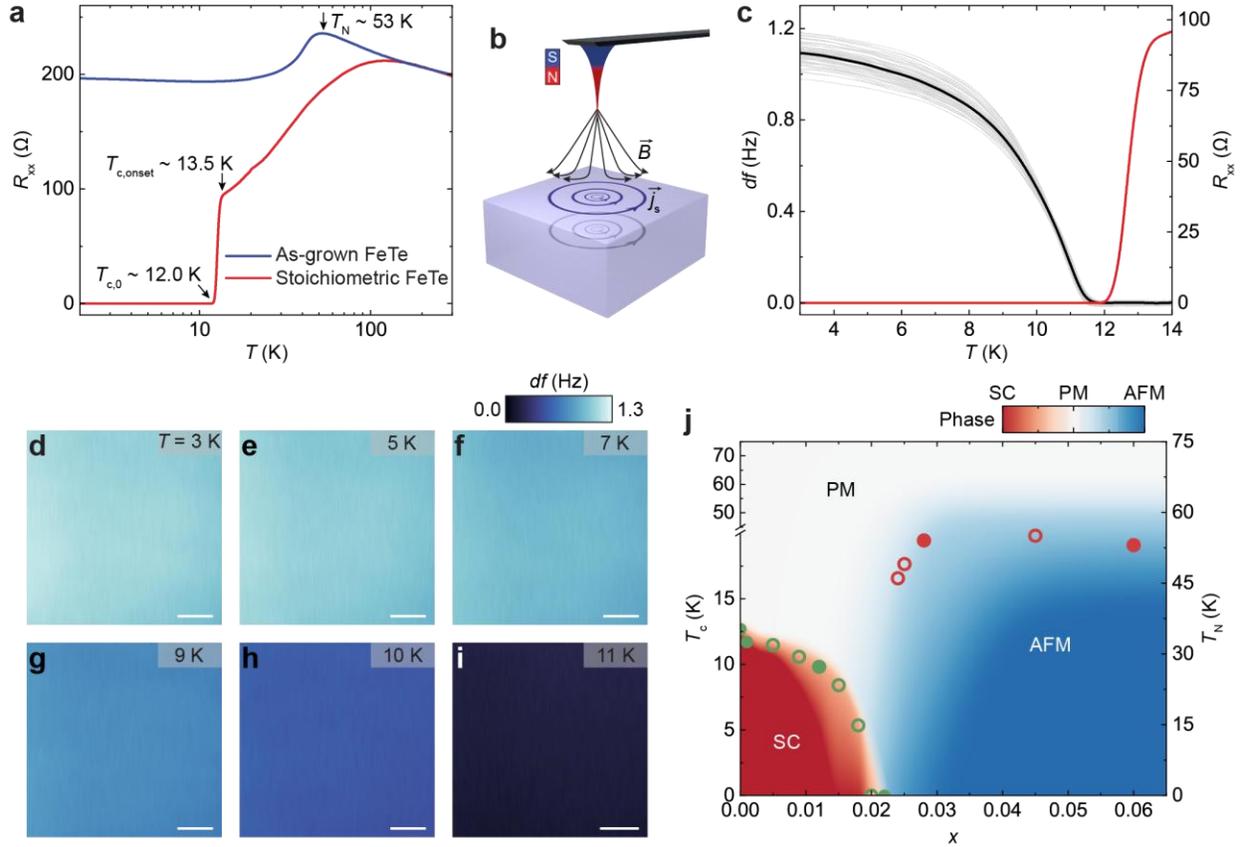

**Fig. 4| Zero-resistance state and Meissner effect in stoichiometric FeTe films. a**, $T$-dependent sheet longitudinal resistance $R_{xx}$ of as-grown (blue) and stoichiometric (red) 40 UC FeTe films. The black arrows indicate the $T_N$ value of the as-grown FeTe film and the $T_{c,onset}$ and $T_{c,0}$ values of the stoichiometric FeTe film. **b**, Schematic of the Meissner effect probed by MFM. Upon approaching the superconducting sample, the magnetic tip experiences a repulsive force due to the screening supercurrents $\vec{J_s}$, which partially expels the stray magnetic field $\vec{B}$ from the tip. **c**, $T$-dependent $R_{xx}$ (red) and MFM frequency shift $df$ (black and gray) of the same stoichiometric 40 UC FeTe film in (**a**), measured over an 8 × 8 grid within an 11 × 11 μm² area. $R_{xx}$ drops nearly to zero at approximately the same $T$ where $df$ begins to increase, confirming that the onset of the Meissner response coincides with the superconducting phase transition. The data are extracted from a series of MFM images measured at different $T$ under zero magnetic field. Each semi-transparent gray line represents the average $df$-$T$ curve from one of the 64 grid cells, and the black line is the overall average. **d-i**, MFM images measured at $T$ = 3 K (**d**), 5 K (**e**), 7 K (**f**), 9 K (**g**), 10 K (**h**), and 11 K (**i**). All MFM images share the same color scale, corresponding to a frequency shift range of 0 to 1.3 Hz. The measurements are performed at a lift height of ~200 nm. Scale bars: 2 μm (**d-i**). **j**, Phase diagram of $Fe_{1+x}Te$, which is plotted as a false-color map. The dark red, dark blue, and white regions denote the superconducting (SC), AFM, and paramagnetic (PM) phases, respectively. The light red and light bule regions represent the SC and AFM phase transition regimes, respectively. In this phase diagram, $T_c$ is defined as the temperature at which $R_{xx}$ drops to 50% of its normal state resistance. Solid circles are obtained from samples treated with integer Te annealing cycles (i.e., as-grown and Cycles I to V), while hollow circles are obtained from samples treated with fractional Te anneal cycles.

**Extended Data Figures:**

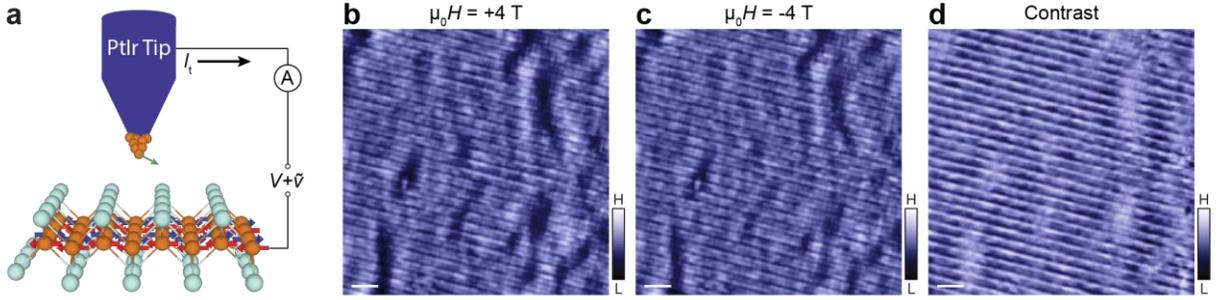

**Extended Data Fig. 1| Spin-polarized STM measurements on an as-grown 40 UC FeTe film under opposite magnetic fields. a**, Schematic of spin-polarized STM measurements on FeTe films. Some iron atoms from the surface of FeTe attached to the apex of the PtIr tip make it spin-polarized. Its polarization direction can be controlled by applying an external magnetic field $\mu_0 H$. **b, c,** STM images (20 × 20 nm$^2$) of the same region in Fig. 1a on the as-grown 40 UC FeTe film at $\mu_0 H = +4$ T (**b**) and $\mu_0 H = -4$ T (**c**) ($V_s = -15$ mV, $I_s = 5$ nA, and $T = 4.2$ K). The STM images in (**b**, **c**) are measured using magnetic Fe-functionalized tips from the surface of FeTe. Under opposite magnetic fields, the spin polarization of the STM tip is reversed, causing the double-stripe pattern to shift by a half-period ($a_{Te}$). **d**, Magnetic contrast image, obtained by subtracting (**b**) from (**c**), which highlights the bicollinear AFM order of the as-grown 40 UC FeTe film.

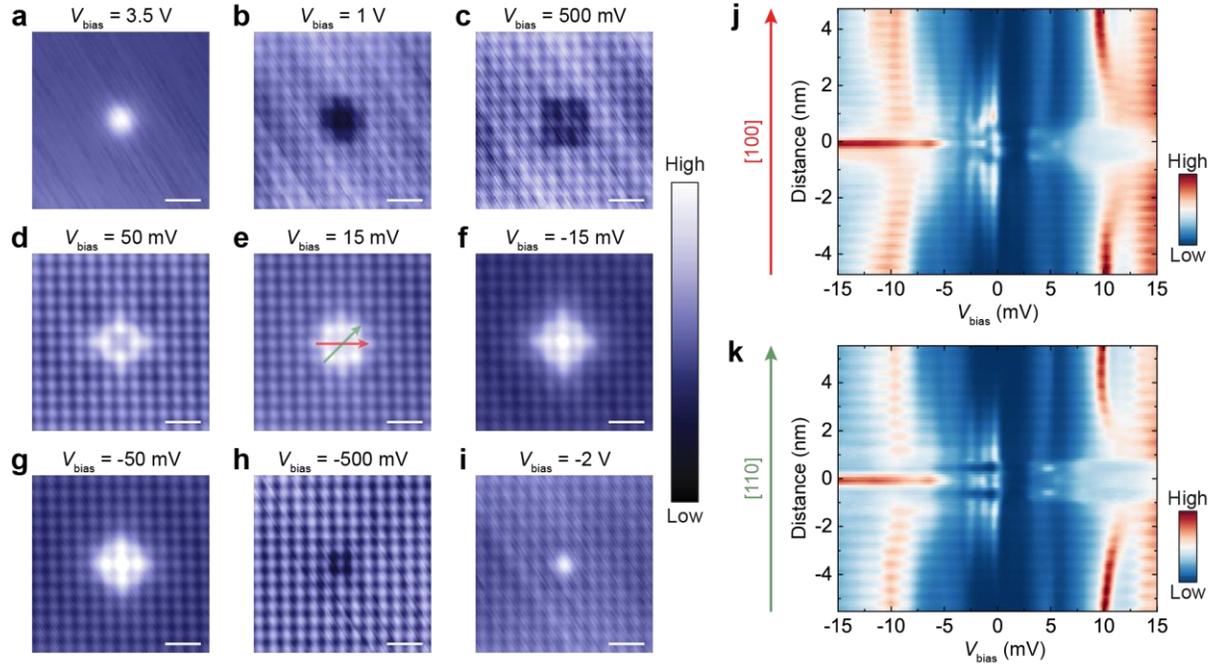

**Extended Data Fig. 2| STM/S results of a single interstitial Fe atom in FeTe films. a-i,** STM images (5 × 5 nm$^2$) of the 40 UC FeTe film with a single interstitial Fe atom measured under different $V_s$. **a**, $V_s$ = 3.5 V and $I_s$ = 1 nA. **b**, $V_s$ = 1 V and $I_s$ = 9 nA. **c**, $V_s$ = 500 mV and $I_s$ = 9 nA. **d**, $V_s$ = 50 mV and $I_s$ = 5 nA. **e**, $V_s$ = 15 mV and $I_s$ = 1 nA. **f**, $V_s$ = −15 mV and $I_s$ = 1 nA. **g**, $V_s$ = −50 mV and $I_s$ = 5 nA. **h**, $V_s$ = −500 mV and $I_s$ = 5 nA. **i**, $V_s$ = −2 V and $I_s$ = 9 nA. **j**, Color plot of d$I$/d$V$ spectra along the red arrow direction in (**e**) (i.e., the [100] direction) across a single interstitial Fe atom ($V_s$ = 15 mV, $I_s$ = 2 nA, and $V_{mod}$ = 0.1 mV). **k**, Color plot of d$I$/d$V$ spectra along the green arrow direction in (**e**) {i.e., the [110] direction} across a single interstitial Fe atom ($V_s$ = 15 mV, $I_s$ = 2 nA, and $V_{mod}$ = 0.1 mV). Several impurity states are resolved near the interstitial Fe atom. Scale bars: 1 nm (**a-i**). The STM images in (**a-i**) are acquired using a spin-polarized tip at $T$ = 4.2 K, while the d$I$/d$V$ spectra are obtained using a PtIr tip at $T$ = 310 mK.

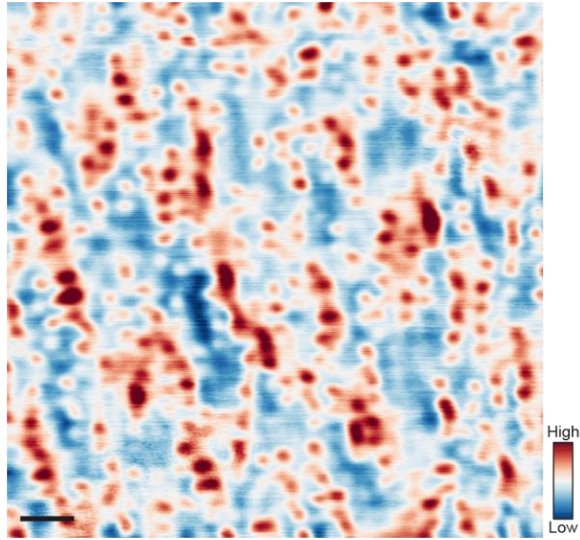

**Extended Data Fig. 3|Distribution of interstitial Fe atoms on the as-grown 40 UC FeTe film.** STM image (20 × 20 nm$^2$) of the same region in Fig. 1a on the as-grown 40 UC FeTe film ($V_s$ = 3.5 V and $I_s$ = 5 nA). The STM image is acquired using a spin-polarized tip at $T$ = 4.2 K. Scale bar: 2 nm.

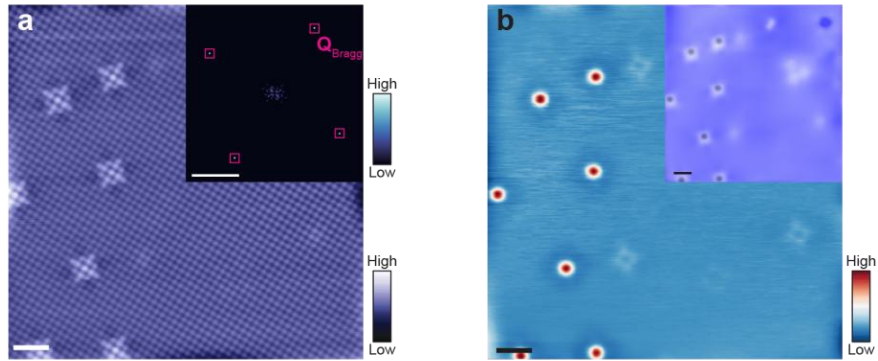

**Extended Data Fig. 4| STM images of the 40 UC FeTe film after 4-cycle Te annealing treatments (i.e., Cycle IV). a**, STM image (20 × 20 nm$^2$) measured using a smaller $V_s$ ($V_s = -10$ mV and $I_s = 1$ nA). Inset: FT image. **b**, STM image (20 × 20 nm$^2$) of the same region in (**a**) measured using a higher $V_s$ ($V_s = 3.5$ V and $I_s = 1$ nA). The red spots in (**b**) represent the interstitial Fe atoms. Inset: The corresponding overlay of the STM images in (**a**) and (**b**). The red (blue) areas in (**b**, inset) represent the AFM (non-AFM) region shown in (**a**). The black spots in (**b**, inset) represent the interstitial Fe atoms in (**b**). The STM images in (**a, b**) are acquired using a spin-polarized tip at $T = 4.2$ K. Scale bars: 2 nm (**a, b**); 1 Å$^{-1}$ (**a**, inset); 2 nm (**b**, inset).

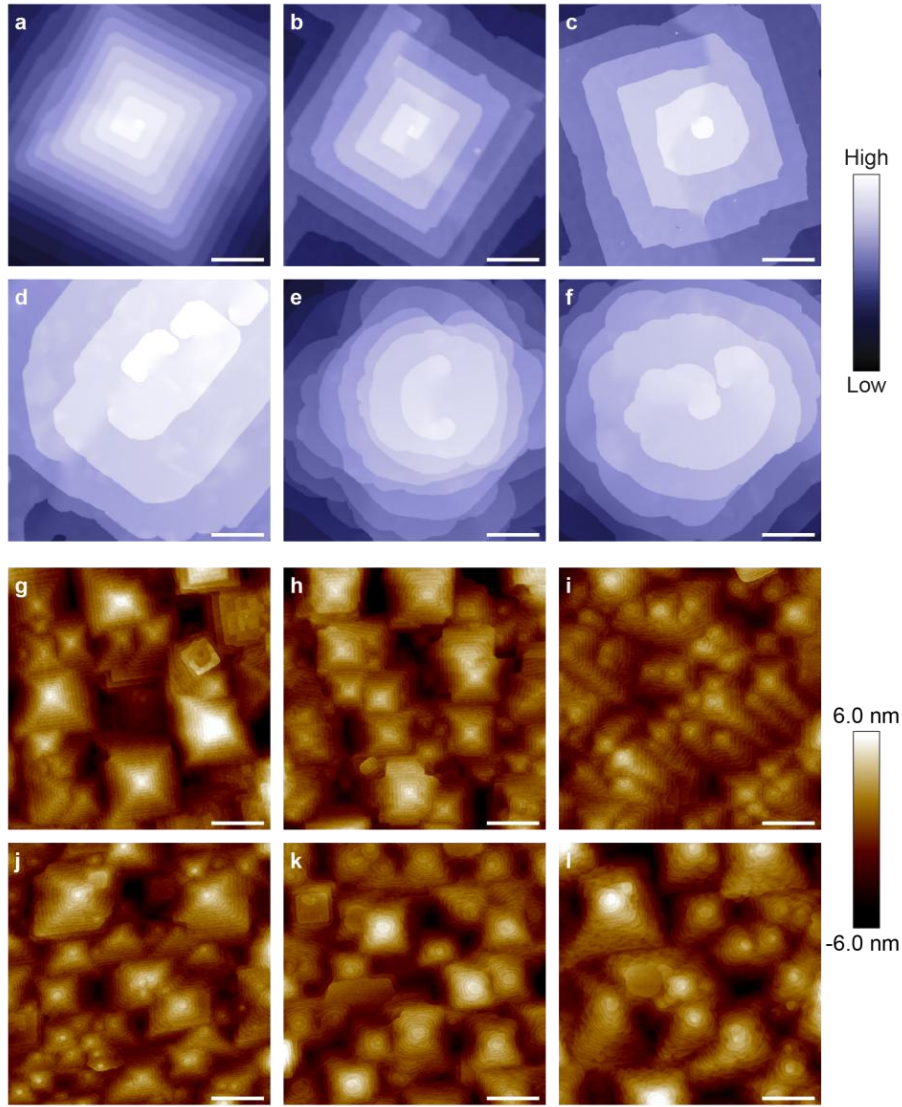

**Extended Data Fig. 5| Surface morphologies of 40 UC FeTe films after each Te annealing cycle. a-f**, STM images (500 × 500 nm$^2$) of 40 UC FeTe films: as-grown FeTe (**a**) ($V_s$ = 1.5 V and $I_s$ = 50 pA), Cycle I (**b**) ($V_s$ = 3.5 V and $I_s$ = 50 pA), Cycle II (**c**) ($V_s$ = 1.5 V and $I_s$ = 20 pA), Cycle III (**d**) ($V_s$ = 1.5 V and $I_s$ = 20 pA), Cycle IV (**e**) ($V_s$ = 1.5 V and $I_s$ = 20 pA), and Cycle V (**f**) ($V_s$ = 1.5 V and $I_s$ = 20 pA). The STM images are acquired using a PtIr tip at $T$ = 4.2 K **g-l**, Atomic force microscopy images (5×5 μm$^2$) of 40 UC FeTe films: as-grown FeTe (**g**), Cycle I (**h**), Cycle II (**i**), Cycle III (**j**), Cycle IV (**k**), and Cycle V (**l**). Scale bars: 100 nm (**a-f**); 1 μm (**g-l**). The as-grown 40UC FeTe film exhibits atomically flat terraces with square pyramidal structures. During the Te annealing treatments, these square terraces gradually expand, and their edges become rounded, suggesting the formation of new FeTe layers.

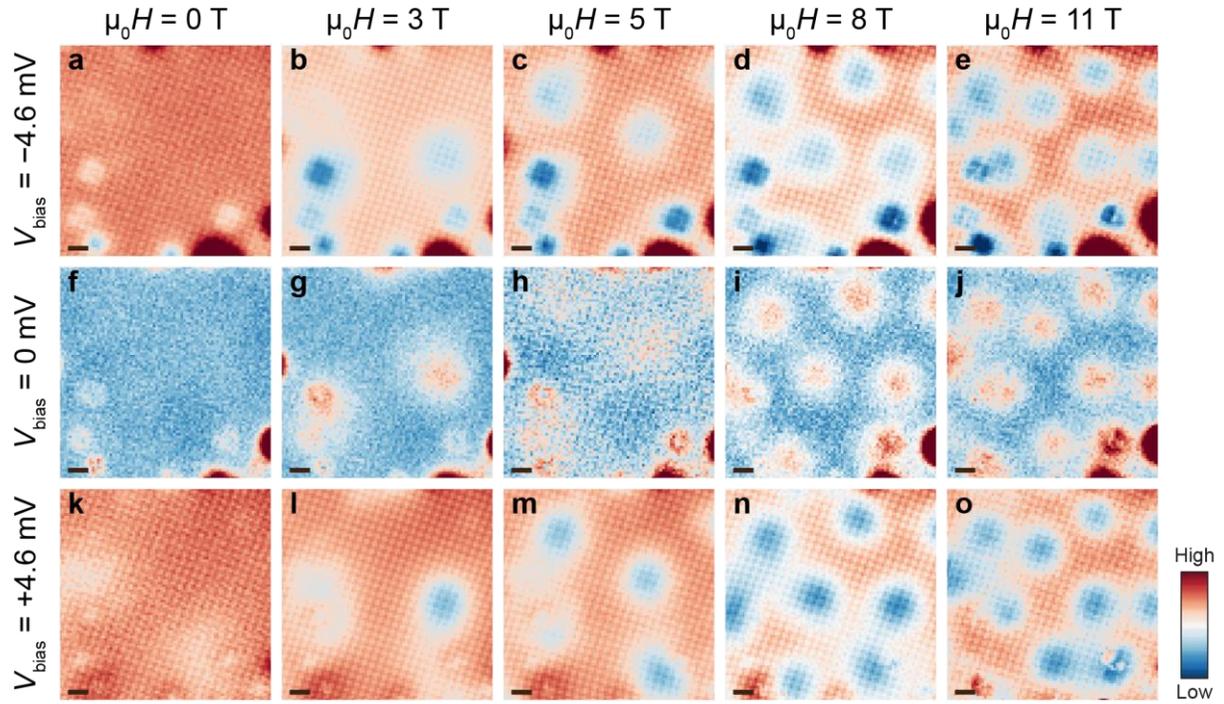

**Extended Data Fig. 6| Abrikosov vortices of a stoichiometric 40 UC FeTe film.** The Abrikosov vortices of a stoichiometric 40 UC FeTe film measured at different $\mu_0 H$ and $V_{bias}$ (42 × 42 nm$^2$, $V_s$ = 15 mV, $I_s$ = 1 nA, and $V_{mod}$ = 2 mV). The image in (**i**) is identical to Fig. 3c. All STM/S measurements are performed using a PtIr tip at $T$ = 310 mK. Scale bars: 4 nm.

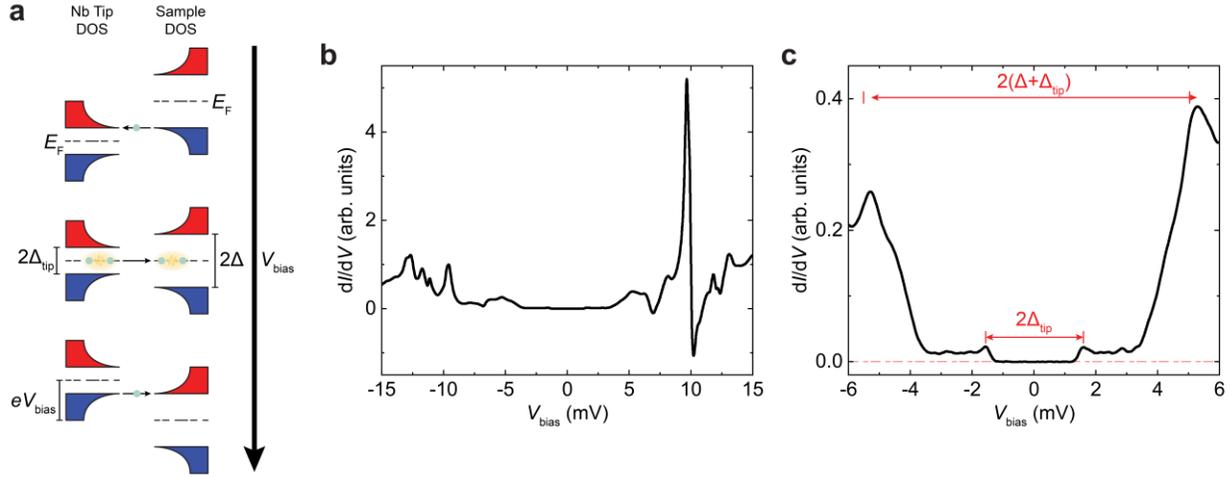

**Extended Data Fig. 7| Josephson STM/S results of a stoichiometric 40 UC FeTe film. a**, Schematics of the tunneling processes at different $V_{bias}$. The blue (red) areas represent the occupied (empty) states, and the dashed lines indicate the chemical potential $E_F$. For $|eV_{bias}| > (\Delta+\Delta_{tip})$, electron tunneling occurs between the tip and the sample. When $eV_{bias}$ is close to zero, Cooper pair tunneling dominates. **b**, Typical d$I$/d$V$ spectrum ($V_s$ = 15 mV, $I_s$ = 2 nA, and excitation voltage $V_{mod}$ = 0.1 mV) of a stoichiometric 40 UC FeTe film measured using a superconducting Nb tip at $T$ = 310 mK with $R_N$ = 7.5 MΩ. **c**, Enlarged d$I$/d$V$ spectrum in (**b**). Two pairs of d$I$/d$V$ peaks have been observed at $\pm\Delta_{tip}$ and $\pm(\Delta+\Delta_{tip})$, where $\Delta_{tip}$ is the superconducting gap size of the Nb tip. See Section I. 3 of Supplementary Information for further discussion regarding non-zero spectral features between $\pm \Delta_{tip}$ and $\pm (\Delta+\Delta_{tip})$.

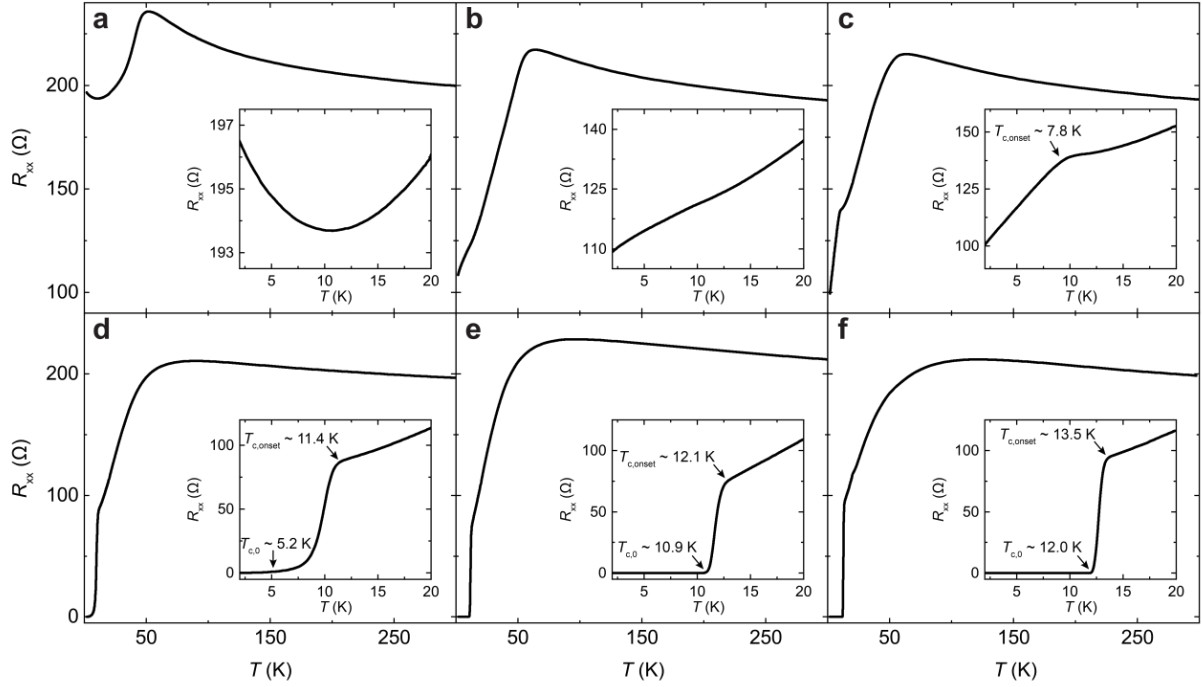

**Extended Data Fig. 8| $R_{xx}$-$T$ curves of 40 UC FeTe films after each Te annealing cycle. a-f**, $T$-dependent $R_{xx}$ of 40 UC FeTe films: as-grown (**a**), after Cycle I (**b**), Cycle II (**c**), Cycle III (**d**), Cycle IV (**e**), and Cycle V (**f**). Inset: Enlarged $R_{xx}$-$T$ curves for 2 K ≤ $T$ ≤ 20 K. The $R_{xx}$-$T$ curves in (**a**) and (**f**) are reused from Fig. 4a.

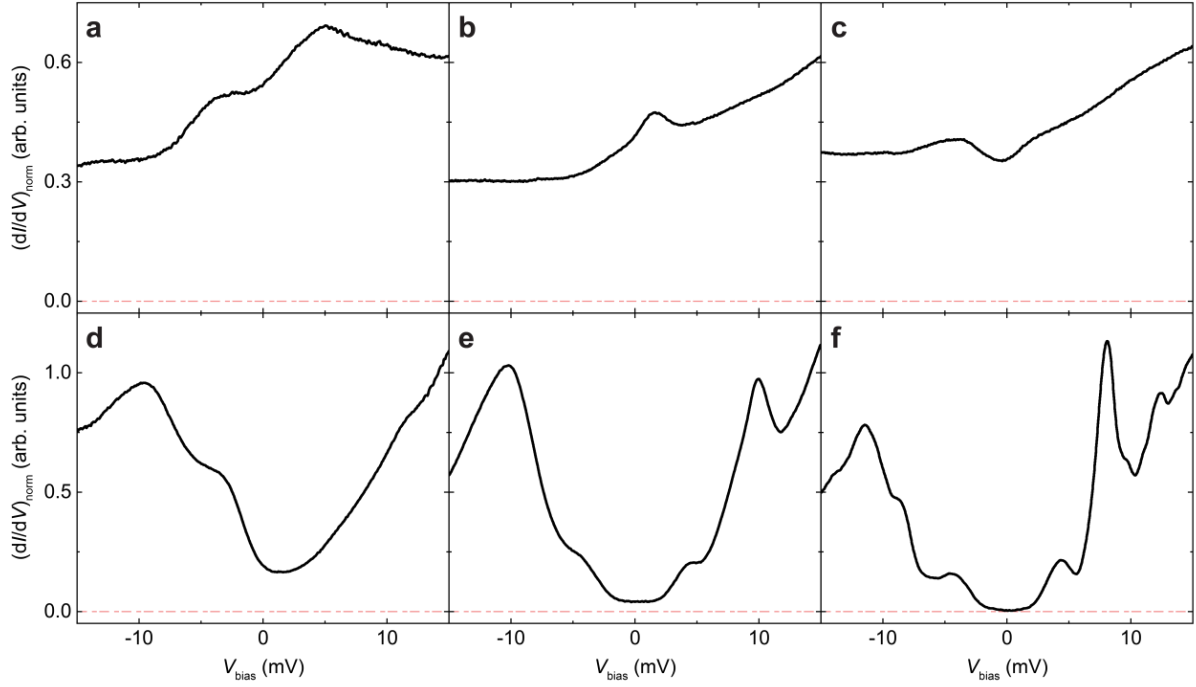

**Extended Data Fig. 9| d$I$/d$V$ spectra of 40 UC FeTe films after each Te annealing cycle. a-f**, Typical d$I$/d$V$ spectra of 40 UC FeTe films: as-grown (**a**) ($V_s$ = 20 mV, $I_s$ = 1 nA, and $V_{mod}$ = 0.2 mV), after Cycle I (**b**) ($V_s$ =15 mV, $I_s$ = 0.3 nA, and $V_{mod}$ = 0.5 mV), Cycle II (**c**) ($V_s$ = 15 mV, $I_s$ = 1 nA, and $V_{mod}$ = 0.2 mV), Cycle III (**d**) ($V_s$ = 15 mV, $I_s$ = 1 nA, and $V_{mod}$ = 0.2 mV), Cycle IV (**e**) ($V_s$ = 15 mV, $I_s$ = 1 nA, and $V_{mod}$ = 0.2 mV), and Cycle V (**f**) ($V_s$ = 15 mV, $I_s$ = 2 nA, and $V_{mod}$ = 0.2 mV). All STM/S measurements are performed using a PtIr tip at $T$ = 4.2 K. To better visualize and compare the d$I$/d$V$ spectra measured under different setpoints, the d$I$/d$V$ spectra in (**a-f**) are normalized by (d$I$/d$V$)$_{norm}$ = (d$I$/d$V$)/($I_s$/$V_s$), which removes the dependence on STM/S setpoint parameters $V_s$ and $I_s$.

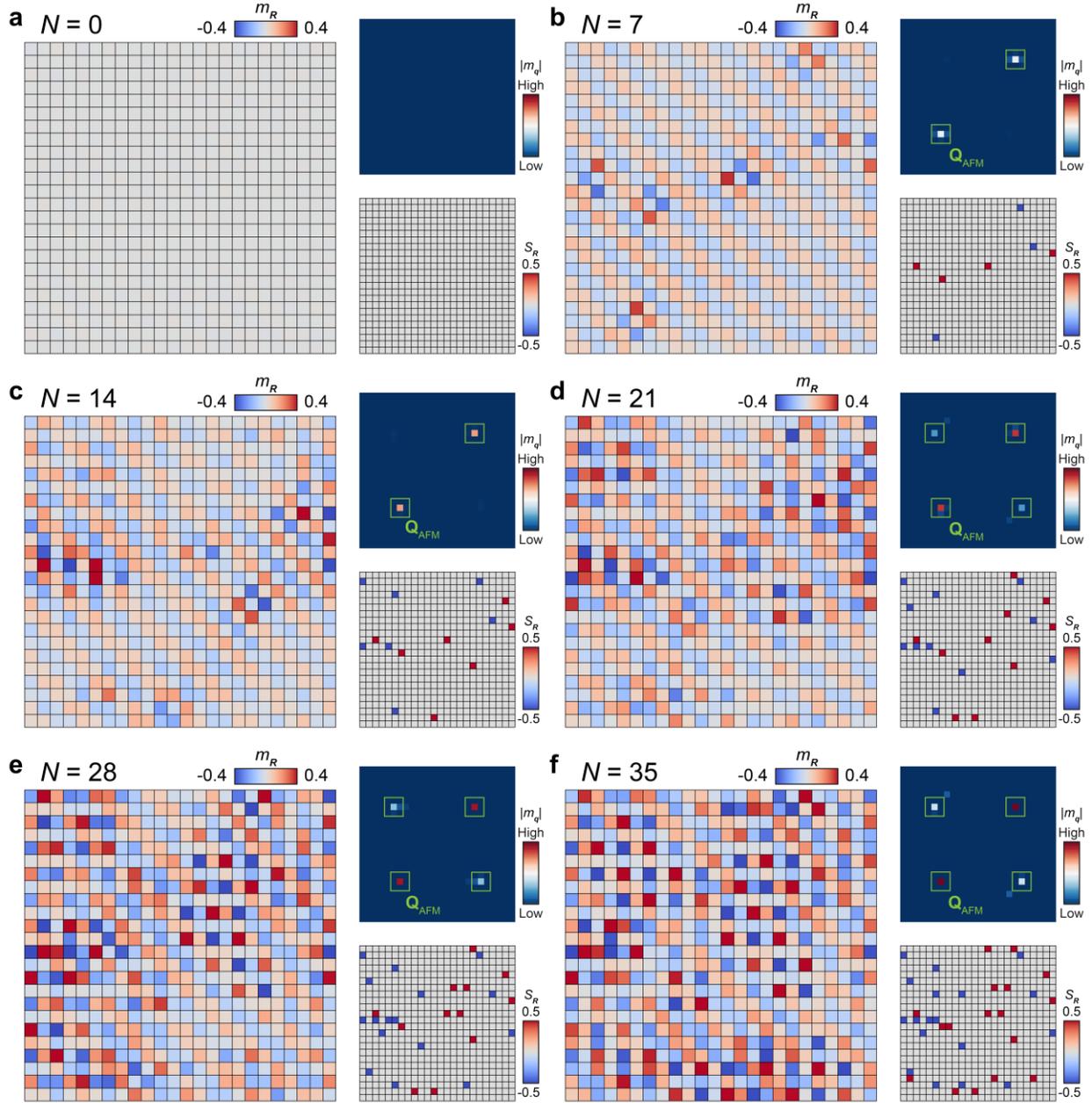

**Extended Data Fig. 10| Numerical calculations on impurity-induced bicollinear AFM order in FeTe. a-e**, Left: Calculated magnetization $m_R$ over 24 × 24 Fe square lattices with $N = 0$ (**a**), $N = 7$ (**b**), $N = 14$ (**c**), $N = 21$ (**d**), $N = 28$ (**e**), and $N = 35$ (**f**). $m_R$ is defined as the difference between the spin-up and spin-down electron probabilities at position $R$. Right top: FT images of $m_R$. Right bottom: Distribution of impurities within the 24 × 24 Fe square lattices, denoted by the magnetic impurity term $S_R$, which is chosen randomly (Methods). All calculations are performed with $V_R = 5$ eV and $|S_R| = 0.5$ eV.